# Supreme Local Lyapunov Exponents and Chaotic Impulsive Synchronization

Sheng Yao Chen, Feng Xi and Zhong Liu

Department of Electronic Engineering

Nanjing University of Science & Technology

Nanjing, Jiangsu 210094

The People's Republic of China

E-mails: chen_shengyao@163.com, xf.njust@gamil.com, eezliu@mail.njust.edu.cn

## Abstract

Impulsively synchronized chaos with criterion from conditional Lyapunov exponent is often interrupted by desynchronized bursts. This is because the Lyapunov exponent cannot characterize local instability of synchronized attractor. To predict the possibility of the local instability, we introduce a concept of supreme local Lyapunov exponent (SLLE), which is defined as *supremum* of local Lyapunov exponents over the attractor. The SLLE is independent of the system trajectories and therefore, can characterize the extreme expansion behavior in all local regions with prescribed finite-time interval. It is shown that the impulsively synchronized chaos can be kept forever if the largest SLLE of error dynamical systems is negative and then the burst behavior will not appear. In addition, the impulsive synchronization with negative SLLE allows large synchronizable impulsive interval, which is significant for applications.

# 1. Introduction

Chaos synchronization is a phenomenon that two or more identical chaotic systems adjust their motions to a common behavior through coupling or forcing. Since the seminal work of Pecora and Carroll [Pecora & Corroll, 1990], the theory and application of the chaos synchronization have acquired wide attention in different research areas. Various synchronization phenomena have been revealed and different synchronization approaches have been reported. See [Parlitz & Kocarev, 1999; Boccaletti *et al*., 2002] for extensive review and references therein.

Impulsive synchronization [Yang & Chua, 1997; Stojanovski & Kocarev, 1997] is a kind of synchronization approaches by which the driven system receives information from the driving system only at discrete time instants. It drastically reduces the amount of synchronization information from the driving system, and is especially suitable to deal with systems which cannot endure continuous disturbances. A lot of the synchronization methods and their applications have been reported [Suykens *et al.*, 1999].

In the study of the impulsive synchronization, one of fundamental issues is the criterion to guarantee the synchronization. The problem is often studied from the error dynamical systems between the coupled systems. The chaos synchronization occurs if and only if the origin of the error system is globally asymptotically stable [Boccaletti *et al*., 2002]. There are two popular approaches to attack the stability. One is to analyze the asymptotical stability of the origin of the error system by Lyapunov stability theory [Chen & Chang, 2009; Sun *et al*., 2002; Wu *et al*., 2007; Yang & Chua, 1997; Yang *et al*., 1997]. If there is a Lyapunov function which asymptotically converges to the origin of the error system, the synchronization is globally asymptotically stable. Another one is to analyze the stability of the linearized error system and results in condition described by the Lyapunov exponents of the error system, which is well-known as conditional Lyapunov exponents (CLEs) [Pecora & Corroll, 1991]. If the largest CLE is negative, the synchronization is *almost* achieved. Since the negative CLE characterizes global stability of synchronization manifold on average, the local instability on the manifold is often overlooked. Therefore, the coupled systems often work in the synchronized state for a long time and then burst into the desynchronized state occasionally (*a.k.a* on-off intermittency [Ott & Sommerer, 1994]). Obviously, the requirement on negative CLE is only necessary condition [Heagy *et al*., 1995], even though it is extensively used in practice as a sufficient condition.

For applications of the impulsive synchronization, large impulsive interval is often required. For example, in areas of chaotic communications/radar [Yang & Chua, 1997; Liu *et al*., 2007] and chaotic compressive sensing [Liu *et al*., 2012], the sampling rates and transmission data can be greatly reduced for the large impulsive interval. However, the above-mentioned synchronization criteria often supply much small impulsive intervals, which are far less than the synchronizable intervals in practice. Even the necessary condition [Heagy *et al*., 1995] can offer large impulsive interval, it does not guarantee the synchronization of high quality [Gauthier & Bienfang, 1996].

This paper tries to derive a sufficient condition for the impulsive synchronization. The proposed condition is based on the observation of the mechanism of the bursting behavior. It is revealed that the bursting is due to the loss of the local transverse stability of the synchronization manifold [Huna *et al*., 1998]. Especially, the bursting behavior takes place in local time with positive local Lyapunov exponent, even when the largest CLE is negative [Muruganandam *et al*., 1999]. In this sense, the burst behavior will not appear if the local instability of the synchronization manifold can be avoided. Therefore, if there is a condition which can avoid the local instability of the synchronization manifold, the synchronization will be completely guaranteed.

To derive the safe synchronization criterion, we introduce in this paper a new Lyapunov exponent, *supreme* local Lyapunov exponent (SLLE), which is defined as supremum of local Lyapunov exponents (LLEs) over the attractor. It is well known that the LLEs characterize the growth or decay rate of the small perturbations to a system trajectory in the prescribed initial point on the attractor and in the prescribed time interval. Then, for chaotic systems, the SLLEs characterize the extreme growth or decay rate of the small perturbations to a system trajectory in the prescribed time interval, regardless of initial points on the attractor. When applied to the impulsively coupled chaotic systems, the synchronized behaviors can be kept forever if the largest SLLE of the error system is negative. This is because the local instability of the error system does not exist and the origin of the error system will be asymptotically stable if the largest SLLE is negative. Application of LLEs to chaos synchronization have been investigated in [Galias, 1999], which gives several criteria through simulations. However, the works in [Galias, 1999] lack rigorous theoretical derivation and are not applicable to the impulsive synchronization.

The criterion of negative SLLE also allows large synchronizable impulsive interval. The negative CLE condition is necessary and its impulsive interval is the largest one. The SLLEs come from Lyapunov exponents and inherit the characteristics of the Lyapunov exponents. Then the impulsive interval derived from the negative SLLE condition is close to that by the negative CLE.

This paper is organized as follows: Section 2 defines the SLLE and analyses its properties. Section 3 derives the sufficient criterion of impulsive synchronization from the SLLEs. Section 4 gives numerical simulations to validate the proposed criterion. Section 5 contains some conclusions.

## 2. Lyapunov Exponents, Local Lyapunov Exponents and Supreme Local Lyapunov Exponents

Consider the following $n$-dimensional continuous-time autonomous chaotic system

$$\dot{\mathbf{x}} = \mathbf{F}(\mathbf{x}) \tag{1}$$

where $\mathbf{x} = [x_1, x_2, ..., x_n]^T \in \mathbf{R}^n$ is the state vector of the system, $\mathbf{F} = [f_1, f_2, ..., f_n]^T \in \mathbf{R}^n$ is the continuous nonlinear vector field of the system and satisfies Lipschitz condition $\|\mathbf{F}(\mathbf{x}) - \mathbf{F}(\mathbf{y})\| \leq L \|\mathbf{x} - \mathbf{y}\|$ with a Lipschitz constant $L$, in which $\|\cdot\|$ denotes the Euclidean norm. Let $\delta \mathbf{x}_0 \in \mathbf{R}^n$ be a small perturbation about initial state $\mathbf{x}_0$ at initial time $t_0$. Then the perturbation dynamics along the trajectory $\mathbf{x}(t)$ obeys the following variational equation

$$\dot{\delta\mathbf{x}}(t) = \mathbf{G}(\mathbf{x}(t))\delta\mathbf{x}(t) \tag{2}$$

where $\mathbf{G}(\mathbf{x}(t))$ is the Jacbian matrix of $\mathbf{F}(\mathbf{x})$ at $\mathbf{x}(t)$. Integrating from $t_0$ to $t_0 + T$, we have

$$\delta\mathbf{x}(t_0 + T) = \mathbf{H}(\mathbf{x}_0, T)\delta\mathbf{x}_0 \tag{3}$$

where $\mathbf{H}(\mathbf{x}_0, T)$ is a linear propagator and depends on the trajectory $\mathbf{x}(t)$ during the finite-time interval $T$. The time evolution of the small perturbations is governed by $\mathbf{H}(\mathbf{x}_0, T)$ forward for the interval $T$.

Let $\mathbf{v}_i(\mathbf{x}_0)$ be the $i$-th right singular vector of $\mathbf{H}(\mathbf{x}_0, T)$ and $\sigma_i(\mathbf{x}_0)$ be the $i$-th singular value. By convention, $\sigma_i(\mathbf{x}_0) \geq \sigma_{i+1}(\mathbf{x}_0)$. The LLEs are defined in [Abarbanel *et al.*, 1991] as

$$\lambda_i^L(\mathbf{x}_0, T) \equiv \frac{1}{T}\ln(\|\mathbf{H}(\mathbf{x}_0, T)\mathbf{v}_i(\mathbf{x}_0)\|) = \frac{\ln(\sigma_i(\mathbf{x}_0))}{T}, \quad i = 1, \cdots, n \tag{4}$$

for the finite-time interval $T$. There are $n$ LLEs for the $n$-dimensional dynamical systems. $\lambda_i^L(\mathbf{x}_0, T)$ characterizes the growth or decay rate of the small perturbation $\delta\mathbf{x}_0$ on the trajectory $\mathbf{x}(t)$ after the finite-time interval $T$ in the direction $\mathbf{v}_i(\mathbf{x}_0)$. The LLEs depend on both the initial state and the finite-time interval.

The LEs are the limits of the LLEs,

$$\lambda_i \equiv \lim_{T \to \infty} \frac{1}{T}\lambda_i^L(\mathbf{x}_0, T) = \lim_{T \to \infty} \frac{1}{T}\ln(\|\mathbf{H}(\mathbf{x}_0, T)\mathbf{v}_i(\mathbf{x}_0)\|), \quad i = 1, \cdots, n \tag{5}$$

According to Oseledec Theorem [Eckmann & Ruelle, 1985], when the finite-time interval $T$ tends to infinity, $\lambda_i^L(\mathbf{x}_0, T)$ converges to $\lambda_i$ for almost every $\mathbf{x}_0$ on the attractor. The LE is global and gives a measure of the average growth or decay rate of the small perturbation $\delta\mathbf{x}_0$ on the attractor. For a strange attractor, there exists at least one positive Lyapunov exponent.

With reference to (4) and (5), we define a concept of supreme local Lyapunov exponents,

$$\lambda_i^S(T) \equiv \sup_{\mathbf{x}_0} \frac{1}{T} \ln\left(\|\mathbf{H}(\mathbf{x}_0,T)\mathbf{v}_i(\mathbf{x}_0)\|\right), \quad i=1,\cdots,n \tag{6}$$

which are the *supremum* of the LLEs over the attractor. It is obvious that the SLLEs characterize the attractor behavior in the finite-time interval. However, different from the LLEs, the SLLEs do not depend on the initial state. In this sense, the SLLEs also contain global characteristics of the attractor. There are $n$ SLLEs with the finite-time interval $T$ for the $n$-dimensional dynamical systems.

The SLLEs are based on the existence of (6). The following proposition confirms the statement.

*Proposition 1:* For a finite-time interval $T$, if the nonlinear vector field of chaotic system satisfies Lipschitz condition $\|\mathbf{F}(\mathbf{x})-\mathbf{F}(\mathbf{y})\| \leq L\|\mathbf{x}-\mathbf{y}\|$, the SLLEs always exist.

*Proof*: With the Lipschitz condition on $\mathbf{F}$, $\|\mathbf{F}(\mathbf{x})-\mathbf{F}(\mathbf{y})\| \leq L\|\mathbf{x}-\mathbf{y}\|$, we know that the Jacbian matrix $\mathbf{G}(\mathbf{x}(t))$ in (2) satisfies

$$\mathbf{G}(\mathbf{x}(t)) \leq L\mathbf{I} \tag{7}$$

where $\mathbf{I}$ is an $n \times n$ identity matrix. Then $L\mathbf{I} - \mathbf{G}(\mathbf{x}(t))$ is positive semi-definite. The $\mathbf{H}(\mathbf{x}_0,T)$ in (3) satisfies

$$\mathbf{H}(\mathbf{x}_0,T) \leq \mathrm{e}^{LT\mathbf{I}} = \mathrm{e}^{LT}\mathbf{I}$$

Then

$$\lambda_i^S(T) = \sup_{\mathbf{x}_0} \frac{1}{T} \ln\left(\|\mathbf{H}(\mathbf{x}_0,T)\mathbf{v}_i(\mathbf{x}_0)\|\right) \leq \frac{1}{T}\ln\left(\mathrm{e}^{LT}\right) = L \tag{8}$$

According to least-upper-bound property [Rudin, 1976], the SLLEs exist. □

It is apparent from (6) that $\lambda_i^S(T) \to \lambda_i \, (i=1,\cdots,n)$ as $T \to \infty$. Moreover, it holds that $\lambda_i^S(T) \geq \lambda_i \, (i=1,\cdots,n)$ for any finite-time interval $T$. This is because

$$\begin{aligned}
\lambda_i &= \lim_{t \to \infty} \frac{1}{t} \ln \left( \| \mathbf{H}(\mathbf{x}_0, t) \mathbf{v}_i(\mathbf{x}_0) \| \right) \\
&= \lim_{N \to \infty} \frac{1}{NT} \ln \left( \| \mathbf{H}(\mathbf{x}(NT-T), NT) \cdots \mathbf{H}(\mathbf{x}_0, T) \mathbf{v}_i(\mathbf{x}_0) \| \right) \\
&\leq \lim_{N \to \infty} \frac{1}{NT} \sum_{k=0}^{N-1} \ln \left( \| \mathbf{H}(\mathbf{x}(kT), (k+1)T) \mathbf{v}_i^k(\mathbf{x}(kT)) \| \right) \\
&\leq \lim_{N \to \infty} \frac{1}{N} \sum_{k=0}^{N-1} \sup_{\mathbf{x}} \frac{1}{T} \ln \left( \| \mathbf{H}(\mathbf{x}, T) \mathbf{v}_i(\mathbf{x}) \| \right) = \lambda_i^S(T)
\end{aligned}$$

Then the SLLEs are always greater than or equal to LLEs and LEs. The SLLEs provide the upper bound of the growth or decay rate of the small perturbations on a trajectory in any finite-time interval $T$ on the attractor. In some application cases, for example, in chaos synchronization and control, it is difficult to locate the system states during the system evolution. The upper bound may be used to find safe conditions for the system stability. Application to chaotic impulsive synchronization is described in next section.

From the definition (6), a direct numerical method to calculate the $i$-th SLLE ($i = 1, \cdots, n$) is to first calculate all $i$-th LLE with the same finite-time interval $T$, regardless of the initial states, and then take the maximum of all calculated $i$-th LLE as the $i$-th SLLE. However, it is impossible to calculate all $i$-th LLE with the same finite-time interval $T$. For the system evolution, we may take the finite-length sliding window to locate the system trajectories with the finite-time interval $T$, as shown in Fig.1. As the evolution going, we can experience all the system trajectories with the finite-time interval $T$ on the attractor. Then the SLLEs are approximately obtained. In the implementation, numerical calculation of the LLEs in [Abarbanel *et al*., 1992] is taken.

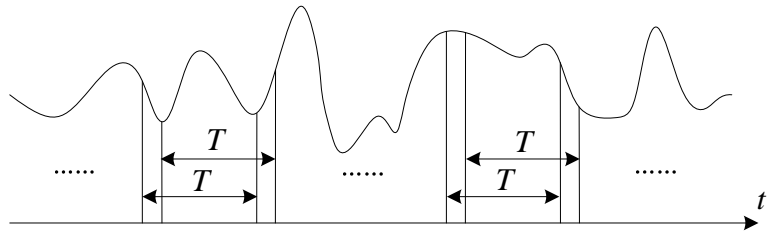

Fig.1. Sliding partitions of the system trajectory.

## 3. SLLEs for Chaotic Impulsive Synchronization

Let us consider an impulsively coupled system with driving system given by (1) and driven system as

$$\begin{cases} \dot{\mathbf{y}} = \mathbf{F}(\mathbf{y}), t \neq t_k^+ \\ \Delta \mathbf{y} \mid_{t=t_k} = \mathbf{y}(t_k^+) - \mathbf{y}(t_k^-) = -\mathbf{Be}, t = t_k^+ \end{cases} \quad (9)$$

where $\mathbf{y} \in \mathbf{R}^n$ is the state vector of driven system, $\mathbf{B}$ is a $n \times n$ impulsive control matrix, $\mathbf{e} = \mathbf{x} - \mathbf{y}$ is the synchronization error, $\mathbf{y}(t_0^+) = \mathbf{y}_0$ and $t_k$ is the impulsive instant. For simplicity, we assume that impulsive interval $\tau$ is equal-distance and $t_k = k\tau$. Then the error system is

$$\begin{cases} \dot{\mathbf{e}} = \mathbf{F}(\mathbf{x}) - \mathbf{F}(\mathbf{y}), & t \neq k\tau \\ \Delta \mathbf{e} \mid_{t=k\tau} = \mathbf{Be}, & t = k\tau \end{cases} \quad (10)$$

In accordance with terms of chaos synchronization, the SLLEs of error system are also called conditional SLLEs (CSLLEs). For $n$-dimensional system, there are $n$ conditional SLLEs. For the impulsive system, the finite-time duration is defined as $[t_0, t_0 + T)$. Then the evolving time of the SLLEs is left continuous, which is consistent with the error system. Different from system (1), the error system (10) is described by impulsive differential equations and it is discontinuous at impulsive instant $t = k\tau$, thus the existence of the SLLEs of error system has to be confirmed again.

*Proposition 2:* For a finite-time interval $T$, if the nonlinear vector field of chaotic system satisfies Lipschitz condition $\|\mathbf{F}(\mathbf{x}) - \mathbf{F}(\mathbf{y})\| \leq L \|\mathbf{x} - \mathbf{y}\|$ and impulsive control matrix satisfies $\mathbf{B} \leq b\mathbf{I}$, where $b > 0$, then the CSLLEs always exist.

*Proof*: Linearizing the error system, we have

$$\begin{cases} \dot{\mathbf{e}} = -\mathbf{G}(\mathbf{x}(t))\mathbf{e}, & t \neq k\tau \\ \Delta \mathbf{e} = \mathbf{Be}, & t = k\tau, k = 1, 2, \cdots \end{cases} \quad (11)$$

With the Lipschitz condition on $\mathbf{F}$, $\|\mathbf{F}(\mathbf{x}) - \mathbf{F}(\mathbf{y})\| \leq L \|\mathbf{x} - \mathbf{y}\|$, we know that the Jacbian matrix $-\mathbf{G}(\mathbf{x}(t))$ in (11) satisfies

$$-\mathbf{G}(\mathbf{x}(t)) \leq L\mathbf{I}$$

Integrating (11) from $t_0$ to $t_0 + T$, we have

$$\mathbf{e}(t_0 + T) = \mathbf{H}(\mathbf{x}_0, \mathbf{y}_0; T)\mathbf{e}_0 \quad (12)$$

where $\mathbf{e}_0 = \mathbf{x}_0 - \mathbf{y}_0$ and $\mathbf{H}(\mathbf{x}_0, \mathbf{y}_0; T)$ is a linear propagator describing the time evolution of (10) for the interval $T$. Then over the range of $[0, t]$ ($t \in (0, \tau)$), we have

$$\mathbf{H}(\mathbf{x}_0, \mathbf{y}_0; t) \leq e^{L t \mathbf{I}} = e^{Lt}\mathbf{I}$$

When $t = \tau$, $\mathbf{H}(\mathbf{x}_0, \mathbf{y}_0; \tau) \leq e^{L\tau}\mathbf{I} + \mathbf{B}$. Let $M = \lfloor T/\tau \rfloor$. Then

$$\mathbf{H}(\mathbf{x}_0, \mathbf{y}_0; M\tau) \leq (e^{L\tau}\mathbf{I} + \mathbf{B})^M$$

and

$$\mathbf{H}(\mathbf{x}_0, \mathbf{y}_0; T) \leq e^{L(T-M\tau)}\mathbf{H}(\mathbf{x}_0, \mathbf{y}_0; M\tau) \leq e^{L(T-M\tau)}(e^{L\tau}\mathbf{I} + \mathbf{B})^M \quad (13)$$

Since the impulsive control matrix satisfies $\mathbf{B} \leq b\mathbf{I}$, we have

$$\mathbf{H}(\mathbf{x}_0, \mathbf{y}_0; T) \leq e^{L(T-M\tau)}(e^{L\tau} + b)^M \mathbf{I} \leq (e^{L\tau} + b)^{M+1}\mathbf{I} \quad (14)$$

Then the CSLLEs satisfy

$$\lambda_i^S(T) = \sup_{\mathbf{x}_0} \frac{1}{T}\ln\left(\|\mathbf{H}(\mathbf{x}_0, \mathbf{y}_0; T)\mathbf{v}_i(\mathbf{x}_0)\|\right) \leq \frac{1}{T}\ln\left((e^{L\tau} + b)^{M+1}\right) = \frac{(M+1)\ln(e^{L\tau} + b)}{T} \quad (15)$$

According to least-upper-bound property [Rudin, 1976], the CSLLEs exist. □

It is well known that the bursting phenomena in the synchronization are due to unstable invariant sets embedded in the synchronization manifold [Heagy *et al.*, 1995]. If the unstable invariant sets are suppressed, the bursting phenomena are avoided. However, it is impractical to locate all the unstable invariant sets in the error system. It has been revealed that the unstable periodic orbits in the unstable invariant sets dominate the instability of the synchronization manifold [Hunt & Ott, 1996a]. In fact, the low period unstable periodic orbits typically achieve the optimal system performance [Hunt & Ott, 1996b]. The mechanism has been used to control chaos and bursting dynamics in chaotic systems [Ott *et al.*, 1990; Nagai *et al.*, 1996]. With the observation, we find that the impulsive synchronization is achieved if the largest CSLLE of the error system (10) is negative for the lowest period of the unstable periodic orbits.

***Theorem 1:*** Let $T_L$ be the lowest period of unstable periodic orbits embedded in the synchronization manifold. If the largest CSLLE of the error system (10) $\lambda_{\max}^S(T_L)$ is negative, the impulsive synchronization is exponentially stable.

***Proof:*** If $\lambda_{\max}^S(T_L) < 0$, the error system (10) will not contain its lowest period unstable periodic orbit. Then for any state $(\mathbf{x}, \mathbf{y})$ and forward finite-time interval $T_L$, we have

$$\mathbf{H}(\mathbf{x}, \mathbf{y}; T_L) \leq e^{\lambda_{\max}^S(T_L) T_L} \mathbf{I} \tag{16}$$

We now show that the synchronization error $\mathbf{e}(t)$ approaches exponentially to zero under the condition $\lambda_{\max}^S(T_L) < 0$.

For any initial state $(\mathbf{x}_0, \mathbf{y}_0)$, let us partition the evolving time of the error system into adjacent time intervals $T_L, 2T_L, \cdots$ and denote the synchronization error as $\mathbf{e}^j(t)$ for the *j-th* time interval $T_L$. Then the $\mathbf{e}^j(t)$ satisfies at $t = jT_L$

$$\mathbf{e}^j(jT_L) = \mathbf{H}(\mathbf{x}((j-1)T_L), \mathbf{y}((j-1)T_L); T_L) \mathbf{e}^{j-1}((j-1)T_L) \leq e^{\lambda_{\max}^S(T_L) T_L} \mathbf{e}^{j-1}((j-1)T_L) \tag{17}$$

which means that the synchronization error satisfies

$$\mathbf{e}^j(jT_L) \leq e^{\lambda_{\max}^S(T_L) jT_L} \mathbf{e}(0)$$

where $\mathbf{e}(0) = \mathbf{x}_0 - \mathbf{y}_0$. When $t \in (jT_L, (j+1)T_L)$, the synchronization error satisfies

$$\mathbf{e}^{j+1}(t) = \mathbf{H}(\mathbf{x}(jT_L), \mathbf{y}(jT_L); t - jT_L) \mathbf{e}^j(jT_L) \leq \mathbf{H}(\mathbf{x}(jT_L), \mathbf{y}(jT_L); t - jT_L) e^{\lambda_{\max}^S(T_L) jT_L} \mathbf{e}(0)$$

According to the Proposition 2, $\mathbf{H}(\mathbf{x}(jT_L), \mathbf{y}(jT_L); t - jT_L)$ has an upper bound

$$\mathbf{H}(\mathbf{x}(jT_L), \mathbf{y}(jT_L); t - jT_L) \leq (e^{L\tau} + b)^{M+1} \mathbf{I} \tag{18}$$

Thus

$$\mathbf{e}^{j+1}(t) \leq (e^{L\tau} + b)^{M+1} e^{\lambda_{\max}^S(T_L) jT_L} \mathbf{e}(0) \tag{19}$$

When $t \to \infty$, $j \to \infty$. For $\lambda_{\max}^S(T_L) < 0$, $\mathbf{e}(t) \to 0$. Then the origin of error system (10) is globally exponentially stable, which means that the impulsive synchronization is globally exponentially stable. □

From (17), the synchronization errors at $t = jT_L$, $\mathbf{e}^j(jT_L)$ ($j = 1, 2, \cdots$), monotonically decreases, if $\lambda_{\max}^S(T_L) < 0$. From (15), the expansion rate of the error system in each interval $T_L$ is not larger than $(M+1)\ln(e^{L\tau} + b)/T_L$. The synchronization error in the $(j+1)$-th interval is confined to a bound, $(e^{L\tau} + b)^{M+1}\mathbf{e}(jT_L)$, which decreases and asymptotically converges to zeros. Therefore, the synchronization error asymptotically converges to zero and the bursting phenomenon cannot appear.

In practical applications, we may need the large impulsive interval for the synchronization. The proposed criterion implicitly addresses the selection of the impulsive interval. However, the criterion is strict and the implied impulsive interval is not much large as simulated in Section 4. It should be noted that the error system trajectories in the time evolution cannot last in the unstable invariant sets successively. It seems that we may calculate the largest CSLLE for large finite-time intervals to check the synchronizability, which will result in large synchronizable impulsive interval. In next section, we will give an empirical method for the selection of the finite-time interval $T$.

The condition by the Theorem 1 does not impose special structure on the impulsive control matrix $\mathbf{B}$. There is much freedom for transmitting the synchronization impulses which may be samples of one or multiple state variables. On the contrary, other existing sufficient conditions often demand that the largest eigenvalue of $(\mathbf{I} + \mathbf{B}^T)(\mathbf{I} + \mathbf{B})$ must be less than 1 [Chen & Chang, 2009; Sun *et al.*, 2002; Wu *et al.*, 2007; Yang & Chua, 1997]. In general, the impulses from samples of all state variables are required. Therefore, the proposed condition can reduce the implementation complexity of impulsive sampling device. On the other hand, the proposed condition allows much larger impulsive interval. This is because the condition is derived from the point of view of the Lyapunov exponents and inherits the characteristics of the CLE necessary condition. However, other sufficient conditions are all derived from the Lyapunov stability and the impulsive intervals are reduced during the derivation by the magnifying or reducing method. Next section will validate these two advantages through simulations.

## 4. Numerical Illustrations

In this section, we take the famous Lorenz system as an example to do the simulation analysis. Three kinds of simulation experiments are conducted. We firstly illustrate the proposed criterion with one state impulse from the driving system and testify its validity to avoid the bursting behavior. Because the criterion is sufficient, we then expose an empirical method to choose the large finite-time interval for CSLLEs with the ability of suppressing the bursting phenomena. Finally, we compare the largest synchronizable impulsive intervals with other sufficient criteria.

The Lorenz system is described by the following differential equations

$$\begin{cases} \dot{x}_1 = p_1(x_2 - x_1) \\ \dot{x}_2 = p_2 x_1 - x_2 - x_1 x_3 \\ \dot{x}_3 = x_1 x_2 - p_3 x_3 \end{cases} \quad (20)$$

where $p_1$, $p_2$ and $p_3$ are the system parameters. For $(p_1, p_2, p_3) = (30, 50, 3)$, (20) works in chaotic state. By the symbolic dynamics method in [Galias & Tucker, 2011], we have calculated the lowest period of the unstable periodic orbits as $T_L = 1.005$.

Let the impulsive control matrix $\mathbf{B} = diag\{[0, -1, 0]\}$. Such a selection of the control matrix means that only one state is sampled and transmitted. Fig.2(a) and Fig.2(b) show the variations of the largest CLE ($\lambda_{\max}$) and the largest CSLLE ($\lambda_{\max}^S(T_L)$) as a function of the impulsive intervals, respectively. It is seen from Fig.2(a) that $\lambda_{\max}$ will be negative for $\tau \leq 0.3075$, which is the necessary condition for the impulsive synchronization. By the proposed criterion, we know from Fig.2(b) that $\lambda_{\max}^{S}(T_L)$ will be negative for $\tau \leq 0.0925$. It is claimed that the bursting behavior will not appear when $\tau \leq 0.0925$.

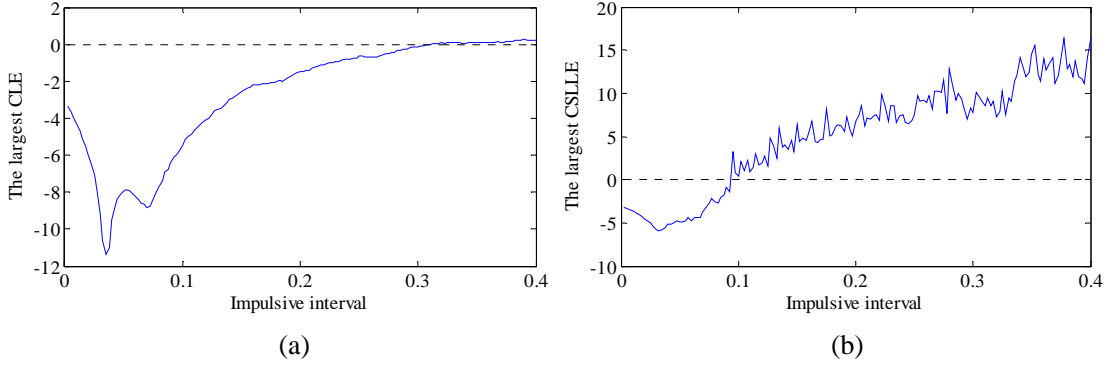

Fig.2. Variation of the largest CLE and the largest CSLLE as a function of the impulsive interval. (a) the largest CLE; (b) the largest CSLLE.

Fig.3 illustrates the probability of occurrence of bursting phenomena as a function of the impulsive intervals. For each impulsive interval, the evolution time of system is confined in 2000 and 500 trials are realized with randomly selected initial states on Lorenz attractor. Impulsive interval is limited to $\tau \leq 0.3075$ to keep the largest CLE negative. As seen in Fig.3, the bursting phenomenon occurs when $0.25 < \tau < 0.26$ and $\tau > 0.28$. The bursting phenomenon is shown in Fig.4(a) and Fig.4(b) for $\tau = 0.255$ ($\lambda_{\max} = -0.6788$, which is far away from zero). The initial states of the driving system and the driven system are $\mathbf{x}_0 = [3.3304, 0.0723, 55.7276]^T$ and $\mathbf{y}_0 = [22.9653, 24.0690, 56.0982]^T$, respectively. Even though synchronization is achieved by a short time in Fig.4(a), the bursting phenomenon occurs after a long time in Fig.4(b). The high-quality synchronization cannot be guaranteed. It is noted that the largest CSLLE $\lambda_{\max}^{S}(T_L)$ is positive at $\tau = 0.255$ as shown in Fig.2(b). Then by the proposed criterion, we may select $\tau = 0.0925$ with negative largest CSLLE $\lambda_{\max}^{S}(T_L)$. The synchronization error is illustrated in Fig.4(c) and Fig.4(d) with the same evolving time and initial states as $\tau = 0.255$. The bursting behavior does not appear. It means that the local instability of the synchronization manifold is avoided and the high-quality synchronization is achieved.

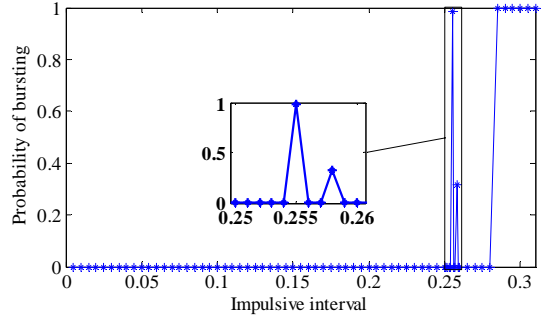

Fig.3. Probability of occurrence of bursting phenomena.

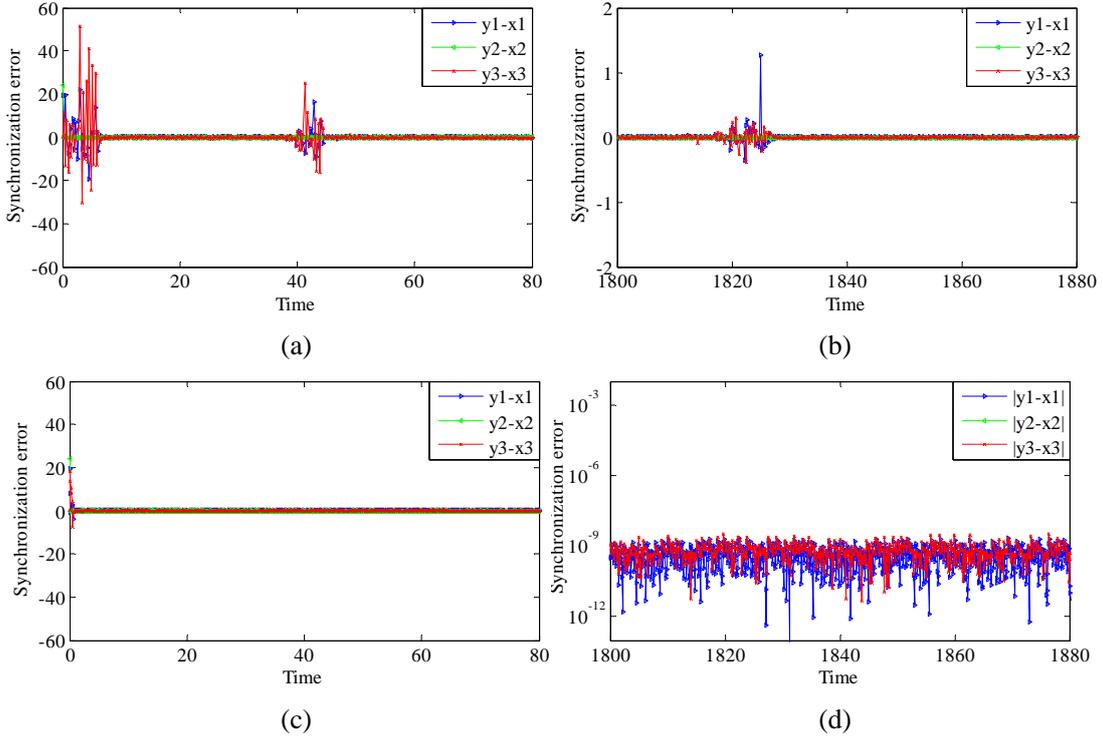

Fig.4. Synchronization error for different impulsive intervals. (a) and (b) $\tau = 0.255$; (c) and (d) $\tau = 0.0925$.

Although the bursting phenomena are suppressed, the largest impulsive interval $\tau_{max} = 0.0925$ is far below that provided by the negative CLE (Fig.2(a)). In practical applications, we may need large impulsive interval. We now expose an empirical method for the selection of the largest impulsive interval. Fig.5(a) shows the variation of the largest CSLLE ($\lambda_{max}^{S}$) as a function of the impulsive intervals for different finite-time intervals of CSLLEs. It is seen that the largest impulsive interval increases as the finite-time interval of CSLLEs increases and falls in the region of negative largest CLE. Fig.5(b) shows the variation of the largest CSLLE as a function of the finite-time intervals $T$ for different impulsive intervals. Obviously, the largest CSLLE monotonously decreases as $T$ increases and changes much slow for different impulsive intervals when $T \geq 50T_L$. Fig.6 illustrates the variation of the largest impulsive interval as a function of the finite-time intervals $T$ of CSLLEs, which indicates that the largest impulsive interval does not change when $T \geq 50T_L$. Therefore, we may choose the finite-time interval of CSLLEs as $T = 50T_L$, which gives the largest impulsive interval as $\tau_{\max} = 0.25$. Fig.7 shows the synchronization error for the same evolving time and initial states in Fig.4. It is seen that the bursting behavior does not appear even for the large impulsive interval.

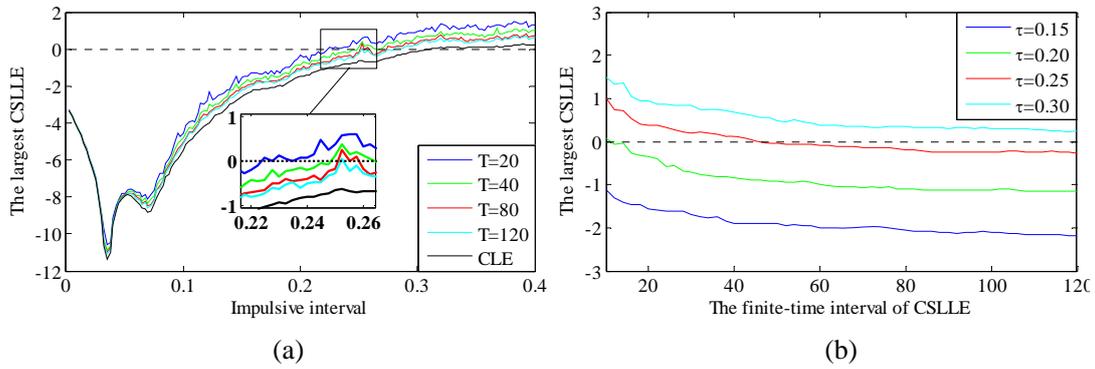

Fig.5. Variations of the largest CSLLE as a function of the impulsive interval and the finite-time interval of CSLLE. (a) the impulsive interval; (b) the finite-time interval of CSLLE.

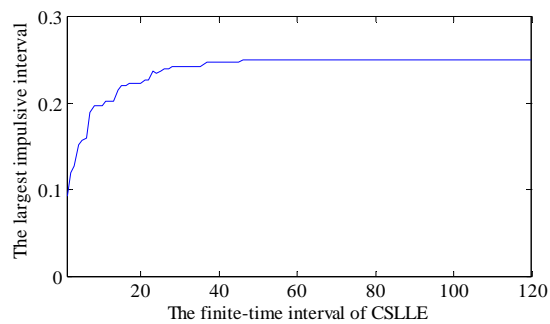

Fig.6. Variation of the largest impulsive interval as a function of the finite-time interval of CSLLE.

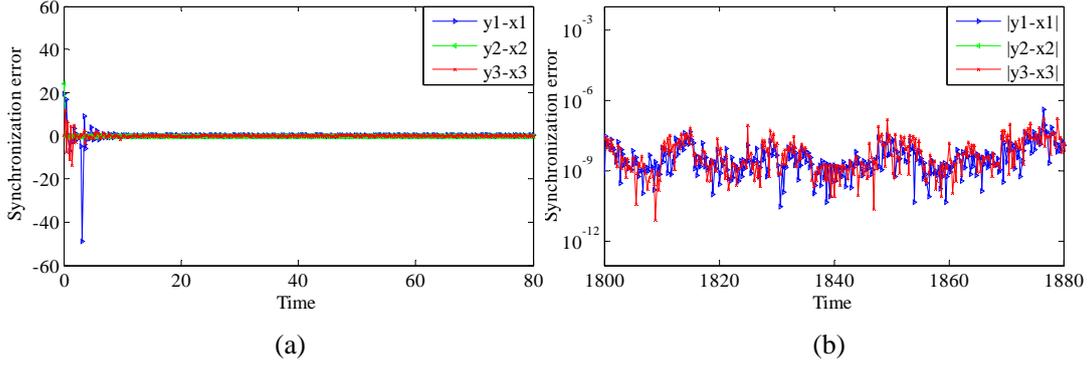

(a)                    (b)

Fig.7. Synchronization error for impulsive interval $\tau = 0.25$.

Finally, the proposed largest synchronizable impulsive interval is compared with other existing sufficient conditions. For clarity, two typical sufficient conditions are discussed here. The first sufficient one is to put forward through a comparison theorem [Yang *et al.*, 1997]. Rewrite the system (1) as

$$\dot{\mathbf{x}} = \mathbf{A}\mathbf{x} + \Phi(\mathbf{x}) \qquad (21)$$

where $\mathbf{A}$ is an $n \times n$ matrix, $\Phi(\mathbf{x}) \in \mathbf{R}^n$ is a continuous nonlinear function vector satisfying the Lipschitz condition $\|\Phi(\mathbf{x}) - \Phi(\mathbf{y})\| \leq L_1 \|\mathbf{x} - \mathbf{y}\|$. Let $d_1$ be the largest eigenvalue of $(\mathbf{I} + \mathbf{B}^T)(\mathbf{I} + \mathbf{B})$ and $q$ be the largest eigenvalue of $(\mathbf{A} + \mathbf{A}^T)$. The origin of the error system is globally asymptotically stable if the equidistant impulsive interval $\tau$ satisfies

$$0 \leq \tau \leq -\frac{\ln(\xi d_1)}{(q + 2L_1)}, \quad \xi > 1 \qquad (22)$$

Another sufficient condition is presented in [Chen & Chang, 2009] with the system (1). For the impulsive synchronization, the equidistant impulsive interval should satisfy

$$0 \leq \tau \leq -\frac{\ln(\xi d_1)}{2L}, \quad \xi > 1 \qquad (23)$$

It is seen that (23) is same as (22) except the denominators at the right hand of inequalities. The difference is resulted from different system descriptions. Obviously, to ensure positive impulsive intervals, $d_1$ must be less than 1 and all state variables of driving system must be used to impulsively control driven system.

For the Lorenz system (20), we have $q = 54.094$, $L_1 = 32.772$ and $L = 45.088$ by numerical simulation. Let control matrix $\mathbf{B} = diag(-0.5, c, -0.5)$ be a diagonal matrix and $-2 \leq c \leq 0$. Fig.8 shows the largest synchronizable impulsive intervals by (22) and (23) and the proposed criterion with $T = T_L$ and $T = 50T_L$. (The sub-figure at the right-upper corner of Fig.8 is the enlarged one by (22) and (23)). For comparison, the largest synchronizable impulsive interval with negative CLE is also given. It is obvious that the largest synchronizable impulsive intervals by the sufficient conditions in [Chen & Chang, 2009; Yang *et al*., 1997] are much smaller than those given by the negative CLE. On the contrary, both the proposed criterion and the empirical selection allow large impulsive intervals, while the impulsive interval by the empirical selection is close to that by the negative CLE.

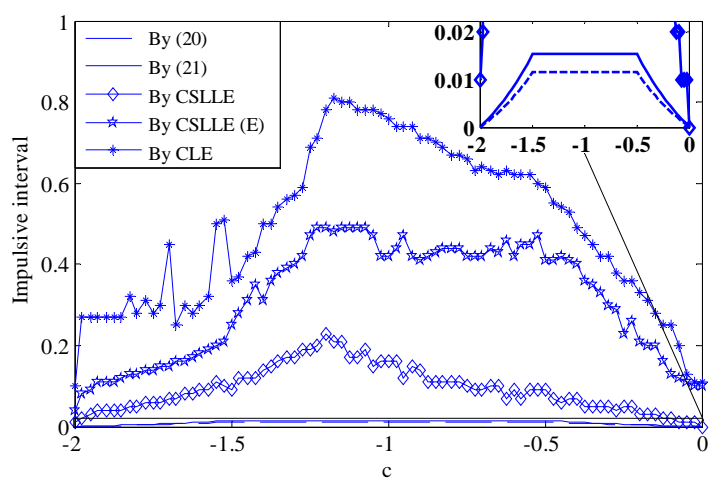

Fig.8. Variations of the largest impulsive interval for different criteria (CSLLE (E) denoting that by the empirical selection).

## 5. Conclusion

Impulsive synchronization is attractive and particularly important to real-life applications. The synchronization criterion from CLE cannot ensure high-quality synchronization. Other sufficient criteria from Lyapunov stability theory do not supply large impulsive intervals to guarantee synchronization. Along the way of the negative CLE condition, this paper proposes a new sufficient criterion for impulsive synchronization with defined SLLE. The SLLE is effective to characterize local instability of the synchronization manifold which often results in the bursting behavior. With negative largest SLLE of the error system, the impulsive synchronization can be kept forever. Simulations show that the proposed condition offers large impulsive intervals over other sufficient conditions.

## Acknowledgments

This work was partially supported by the National Science Foundation of China (60971090, 61171166, 61101193).